\documentclass[12pt]{article}
 \usepackage{epsfig}
 \def\be{\begin{equation}}
 \def\ee{\end{equation}}
 \def\bea{\begin{eqnarray}}
 \def\eea{\end{eqnarray}}
 \usepackage{graphicx}
\usepackage{amssymb}
\usepackage{xcolor}

 \catcode`\@=11
 \def\lsim{\mathrel{\mathpalette\@versim<}}
 \def\gsim{\mathrel{\mathpalette\@versim>}}
 \def\@versim#1#2{\vcenter{\offinterlineskip
 \ialign{$\m@th#1\hfil##\hfil$\crcr#2\crcr\sim\crcr } }}
 \catcode`\@=12

 \catcode`@=12
\begin{document}
 \thispagestyle{empty}
 \begin{flushright}
 UCRHEP-T614\\
 May 2021\
 \end{flushright}
 \vspace{0.6in}
 \begin{center}
 {\LARGE \bf Radiative Seesaw Dark Matter\\}
\vspace{1.2in}
 {\bf Ernest Ma\\}
 \vspace{0.1in}
{\sl Department of Physics and Astronomy,\\ 
University of California, Riverside, California 92521, USA\\}
\vspace{0.2in}
{\bf Valentina De Romeri\\}
{\sl Institut de F\'{i}sica Corpuscular CSIC/Universitat de Val\`{e}ncia, Parc Cient\'ific de Paterna\\
 C/ Catedr\'atico Jos\'e Beltr\'an, 2 E-46980 Paterna (Valencia) - Spain}
 \vspace{0.1in}

\end{center}
\vspace{1.0in}

\begin{abstract}\
The singlet majoron model of seesaw neutrino mass is appended by one dark 
Majorana fermion singlet $\chi$ with $L=2$ and one dark complex scalar 
singlet $\zeta$ with $L=1$.  This simple setup allows $\chi$ to obtain 
a small radiative mass anchored by the same heavy right-handed neutrinos, 
whereas the one-loop decay of the standard-model 
Higgs boson to $\chi \chi + \bar{\chi} \bar{\chi}$ provides the freeze-in 
mechanism for $\chi$ to be the light dark matter of the Universe.

\end{abstract}

\newpage
\baselineskip 24pt

\noindent \underline{\it Introduction}~:~ 
Neutrino mass~\cite{g16} and dark matter~\cite{y17} are two outstanding 
issues in particle physics and astrophysics.  They may be essentially 
related, such as in scotogenic models~\cite{m06} where the former requires 
the existence of the latter.  They may also be indirectly related through 
lepton parity~\cite{m15} or lepton number~\cite{m20}.

Dark matter is conventionally believed to be a massive particle of order 
100 GeV and interacts weakly with matter.  Its annihilation cross section 
$\times$ relative velocity at rest should be about 
$3 \times 10^{-26}~{\rm cm}^3/{\rm s}$, as it 
freezes out of thermal equilibrium in the early Universe.  It should then 
be observable in direct search experiments in underground laboratories.

An alternative is the freeze-in mechanism~\cite{Hall:2009bx} where the dark 
matter interacts very weakly and slowly builds up its relic abundance, from 
the decay of a massive particle for example before the latter itself goes 
out of thermal equilibrium.  In this case, the direct detection of dark matter 
in underground experiments becomes very difficult, which is consistent 
with the mostly null results obtained so far.

The standard model (SM) of particle interactions conserves baryon number $B$ 
and $L$ automatically.  If a singlet right-handed fermion $N_R$ is added, 
the term $\bar{N}_R (\nu_L \phi^0 - e_L \phi^+)$ is allowed by gauge 
invariance, hence $N_R$ is naturally assigned $L=1$.  On the other hand, 
gauge invariance also allows the $N_R N_R$ Majorana mass term, hence $L$ 
naturally breaks to $(-1)^L$ and a seesaw mass for $\nu_L$ is obtained 
as is well-known.  Instead of letting gauge invariance decide on all 
the global and discrete symmetries of the Lagrangian, the latter may be 
imposed as additional inputs.  In that case, other choices for $N_R$ 
are also possible~\cite{m17}.

Returning to the canonical choice of $L=1$ for $N_R$, if a new complex 
neutral scalar singlet $\eta$ with $L=2$ is added, then the term 
$\eta^* N_R N_R$ is allowed and the Lagrangian with this term replacing 
the $N_R N_R$ term is invariant under the global symmetry $U(1)_L$. 
This conscious choice of a new particle with nonzero lepton number was 
first made 40 years ago~\cite{cmp81}, where $\eta$ is assumed to have a nonzero 
vacuum expectation value, thereby breaking $U(1)_L$ spontaneously again 
to $(-1)^L$, but also with the appearance of a massless Goldstone boson 
called the singlet majoron.  The seesaw mechanism applies as before.

In this paper, the conscious choice of two additional particles with 
nonzero lepton number is made.  $\zeta$ is a neutral complex scalar 
with $L=1$ and $\chi_L$ is a neutral Majorana fermion with $L=2$. 
As expected~\cite{m15} from $(-1)^{L+2j}$, they are particles of odd dark 
parity.  It will be shown that $\chi_L$ obtains a small radiative mass 
anchored by $N_R$ and it becomes dark matter through the freeze-in mechanism 
from Higgs decay.

\noindent \underline{\it Description of Model}~:~ 
The SM is extended first by three singlet right-handed neutrinos $N_R$ with 
lepton number $L=1$ as well as one complex scalar $\eta$ with $L=2$, so 
that the term $\eta^* N_R N_R$ appears in the Lagrangian, which is assumed 
invariant under $U(1)_L$.  However, $\langle \eta \rangle \neq 0$ breaks 
$U(1)_L$ spontaneously, resulting in a singlet majoron~\cite{cmp81}.
This well-known model is now extended to include one complex scalar $\zeta$ 
with $L=1$, and one Majorana fermion $\chi_L$ with $L=2$.  The 
Lagrangian of the SM is then expanded, having 
\begin{eqnarray}
{\cal L}_Y &=&   - {1 \over 2} f_N \eta^* N_R N_R - f_\nu \bar{N}_R 
(\nu_L \phi^0 - l_L \phi^+) - f_\chi \bar{\chi}_L N_R \zeta + H.c.,
\end{eqnarray}
as well as
\begin{eqnarray}
V(\zeta, \eta, \Phi) &=& -\mu_0^2 \Phi^\dagger \Phi - \mu_2^2 \eta^* \eta 
 + m_1^2 \zeta^*\zeta + {1 \over 2} \mu_1 \eta^* \zeta^2 + H.c. \nonumber \\ 
&& +~{1 \over 2} \lambda_0 (\Phi^\dagger \Phi)^2 + {1 \over 2} \lambda_1 
(\zeta^*\zeta)^2 + {1 \over 2} \lambda_2 (\eta^*\eta)^2 \nonumber \\  
&& +~\lambda_{01} (\zeta^* \zeta)(\Phi^\dagger \Phi) + 
\lambda_{02} (\eta^* \eta)(\Phi^\dagger \Phi) + 
\lambda_{12} (\zeta^* \zeta)(\eta^* \eta).
\end{eqnarray}
In the above, lepton number $L$ is conserved, but is spontaneously broken to 
lepton parity $(-1)^L$ by $\langle \eta \rangle$.  Note that the $f_N$ term 
justifies the assignment of $L=2$ to $\eta$, and the $\mu_1$ term justifies 
the assignment of $L=1$ to $\zeta$, and the $f_\chi$ term justifies the 
assignment of $L=2$ to $\chi_L$.  Without $\eta$, the separate assignments 
of $L$ to $\zeta$ and $\chi_L$ would have been ambiguous.  Let
\begin{equation}
\Phi = \pmatrix{\phi^+ \cr \phi^0} =  
\pmatrix{0 \cr (v+h)/\sqrt{2}}, ~~~ \eta = \left( u + {\rho \over \sqrt{2}} 
\right) e^{i\theta/u\sqrt{2}}, ~~~ \zeta = {1 \over \sqrt{2}} 
(\zeta_R + i \zeta_I) e^{i\theta/2u\sqrt{2}},
\end{equation}
then $\eta^* \zeta^2 + H.c. = (u+\rho/\sqrt{2})(\zeta_R^2-\zeta_I^2)$. The 
minimum of $V$ is determined by
\begin{eqnarray}
\mu_0^2 &=& {1 \over 2} \lambda_0 v^2 + \lambda_{02} u^2, \\ 
\mu_2^2 &=& \lambda_2 u^2 + {1 \over 2} \lambda_{02} v^2.
\end{eqnarray}
As a result, the mass-squared matrix spanning $h$ and $\rho$ is given by
\begin{equation}
{\cal M}^2_{h \rho} = \pmatrix{ \lambda_0 v^2 & \sqrt{2} \lambda_{02} vu \cr 
\sqrt{2} \lambda_{02} vu & 2 \lambda_2 u^2}.
\end{equation}
Now the $N_R$ mass is $m_N = f_N u$, hence $u$ should be much greater than 
$v$.  This means that $h$ is predominantly the SM Higgs boson and mixes 
very little with the very heavy $\rho$.  As for the $\zeta_{R,I}$ masses, 
they are split by the $\mu_1$ term, i.e.
\begin {eqnarray}
m_R^2 &=& m_1^2 + {1 \over 2} \lambda_{01} v^2 + \lambda_{12} u^2 + \mu_1 u, \\
m_I^2 &=& m_1^2 + {1 \over 2} \lambda_{01} v^2 + \lambda_{12} u^2 - \mu_1 u.
\end{eqnarray}
The neutrinos obtain canonical seesaw masses $m_\nu = m_D^2/m_N$, where 
$m_D = f_\nu v/\sqrt{2}$.

\noindent \underline{\it Light Dark Majorana Fermion}~:~ 
The $\bar{\chi}_L N_R \zeta$ term is similar to the one found in the prototype 
renormalizable model~\cite{prv08} of fermionic dark matter $\chi_L$ 
supplemented by a real scalar counterpart $S$, i.e. $\bar{\chi}_L N_R S$.  
In that case, odd dark parity is imposed on $\chi$ and $S$.  However,  
if lepton parity $L_P$ is used, i.e. even for $\chi$ and odd for $S,N$, 
the same dark parity $D_P$ is derived~\cite{m15}, namely 
$D_P = L_P (-1)^{2j}$.  Note that $\chi$ is allowed an arbitrary Majorana 
mass.  This scenario admits either $\chi$ or $S$ to be dark matter, and 
has been studied extensively.

In the present model, $L$ is conserved in the Lagrangian, hence $\chi$ (with 
$L=2$) is massless at tree level.  After $L$ is spontaneously broken by two 
units, $\chi$ then obtains a radiative seesaw mass as shown in Fig.~1.  
Its structure is analogous to that of the scotogenic model~\cite{m06}, with 
its finiteness coming from the cancellation of the $\zeta_{R,I}$ 
contributions.  Both $\chi$ and $\zeta_{R,I}$ belong to the dark sector, 
with odd dark parity $D_P$.
\begin{figure}[htb]
\vspace*{-6cm}
\hspace*{-3cm}
\includegraphics[scale=1.0]{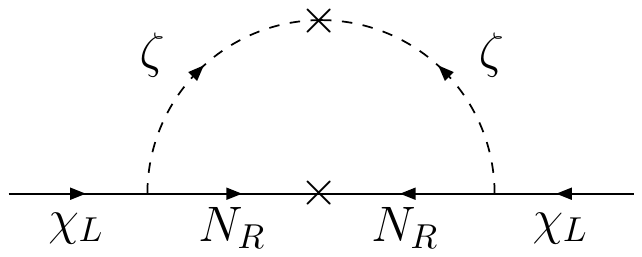}
\vspace*{-21.5cm}
\caption{One-loop radiative Majorana mass for the dark fermion $\chi$.}
\end{figure}
The one-loop diagram of Fig.~1 is easily calculated. 
\begin{equation}
m_\chi = {f_\chi^2 m_N \over 32 \pi^2} \left[ 
{m_R^2 \ln (m_R^2/m_N^2) \over m_R^2-m_N^2} - 
{m_I^2 \ln (m_I^2/m_N^2) \over m_I^2-m_N^2} \right].
\end{equation}
For $m_R^2-m_I^2=2\mu_1 u<<m_\zeta^2=(m_R^2+m_I^2)/2<<m_N^2$, this becomes
\begin{equation}
m_\chi ={f_\chi^2 \mu_1 u\over 16 \pi^2 m_N} \left[ \ln {m_N^2 \over m_\zeta^2} 
- 1 \right].
\end{equation} 
Note that both $m_\nu$ and $m_\chi$ are of the seesaw form, anchored 
by $m_N$.  A tree-level seesaw for light fermion dark matter is also 
possible in the context of $U(1)_\chi$~\cite{m18,m19}.

\noindent \underline{\it Production of Light Fermion Dark Matter}~:~
The only interaction involving $\chi$ is the $f_\chi \bar{\chi}_L N_R \zeta$ 
term.  Assuming that the reheat temperature of the Universe is low enough 
(say a few TeV), its relic abundance comes from freeze-in through Higgs 
decay, with the effective interaction  
\begin{equation}
{\cal L}_{int} = {1 \over 2} f_h h (\chi \chi + \bar{\chi}\bar{\chi}),
\end{equation}
where $f_h$ is given by Fig.~2.
\begin{figure}[htb]
\vspace*{-5cm}
\hspace*{-3cm}
\includegraphics[scale=1.0]{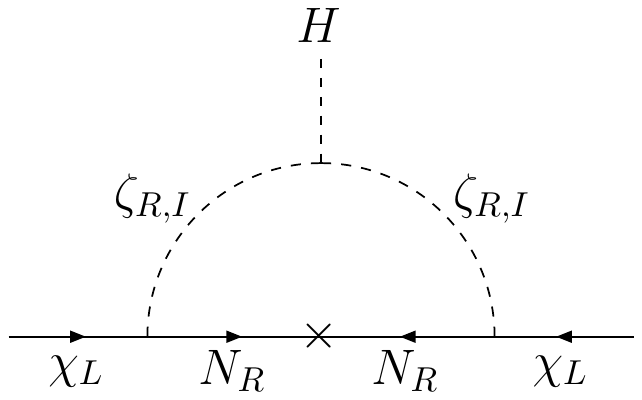}
\vspace*{-21.5cm}
\caption{Direct one-loop diagram for Higgs decay to $\chi \chi$.}
\end{figure}
\begin{eqnarray}
f_h &=& {\lambda_{01} v f_\chi^2 m_N \over 32 \pi^2} \left[ 
{1 \over m_R^2-m_N^2} - {m_N^2 \ln(m_R^2/m_N^2) \over (m_R^2-m_N^2)^2} - 
{1 \over m_I^2-m_N^2} + {m_N^2 \ln(m_I^2/m_N^2) \over (m_I^2-m_N^2)^2}  
\right] \nonumber \\ 
&=& {-\lambda_{01} v f_\chi^2 \mu_1 u\over 16 \pi^2 m_N m_\zeta^2} ~=~ 
{-\lambda_{01} v m_\chi \over m_\zeta^2} \left[ \ln {m_N^2 \over m_\zeta^2} - 1 
\right]^{-1}.
\end{eqnarray}
The decay rate of $h$ to $\chi \chi + \bar{\chi}\bar{\chi}$ is 
\begin{equation}
\Gamma_h = {f_h^2 m_h \over 32\pi} \sqrt{1-4x^2}(1-2x^2),
\end{equation}
where $x=m_\chi/m_h$.

Another one-loop diagram for $h$ decay to $\chi \chi$ is shown in Fig.~3.
\begin{figure}[htb]
\vspace*{-5cm}
\hspace*{-3cm}
\includegraphics[scale=1.0]{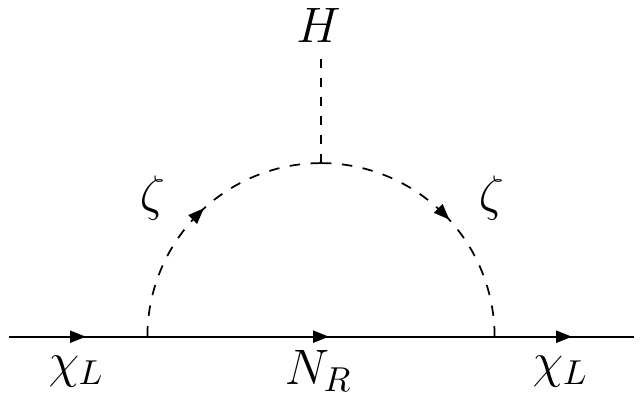}
\vspace*{-21.5cm}
\caption{One-loop diagram for Higgs decay to $\chi \chi$ through external 
$m_\chi$ insertion.}
\end{figure}
The corresponding effective coupling is given by
\begin{equation}
f_h' = {\lambda_{01} v m_\chi f_\chi^2 \over 16 \pi^2 m_N^2} \left[ 
\ln {m_N^2 \over m_\zeta^2} - {3 \over 2} \right].
\end{equation}
Hence $|f_h'| << |f_h|$ for $m_\zeta << m_N$.


\noindent \underline{\it Phenomenology of Fermionic Dark Matter}~:~
We assume that the reheat temperature of the Universe after inflation is 
below the decoupling temperature of $\chi$ but above $m_H$, say 
$T_R \sim 1-10$ TeV.  In such a scenario, $\chi$ is a feebly interacting 
massive particle (FIMP) which only production mechanism is 
freeze-in~\cite{Hall:2009bx,McDonald:2001vt} through Higgs decay, before 
the latter decouples from the thermal bath.  (In principle, there could also 
be $2\rightarrow 2$ scattering of $\chi$ with other states in the thermal 
bath, mediated by the Higgs boson~\cite{Arcadi:2013aba,Blennow:2013jba}.) 
The relic abundance of $\chi$, initially negligible in the early Universe, 
gradually grows via its feeble coupling $f_h$.  Both $N_R$ and $\zeta$ are 
taken to be much heavier than $T_R$ so that their abundances are Boltzmann 
suppressed. On the contrary, if $m_N$ or $m_{\zeta}$ $\lesssim T_R$, new 
interaction channels would open and the relic density of $\chi$ could 
excessively increase due to the term 
$f_\chi \bar{\chi}_L N_R \zeta$, such as  
$N_R \bar{N_R} \rightarrow \chi \bar{\chi}$, 
$\zeta \zeta \rightarrow \chi \chi$, or the decays of $\zeta$. 
These processes could possibly lead to the thermalization of $\chi$.  
Even if $m_N (m_{\zeta}) >> T_R$, the channel 
$\nu \bar{\nu} \rightarrow \chi \bar{\chi}$ is always present, although it is 
suppressed by the mixing of $N_R$ with the light active neutrinos. Eventually, 
we find that this channel never contributes sizeably to the thermalization of 
$\chi$.

We perform a scan varying randomly the main free parameters which characterize 
the model and we rely on public computer tools to 1) implement the model 
with ({\tt FeynRules 2.0}~\cite{Alloul:2013bka} and 
{\tt CalcHEP 3.4}~\cite{Belyaev:2012qa}), and 2) numerically compute the 
relic abundance of $\chi$ via freeze-in with 
({\tt Micromegas 5.2.4}~\cite{Belanger:2018ccd}). In the analysis we require 
the parameters to comply with the hierarchy of scales 
$2\mu_1 u<<m_\zeta^2<<m_N^2$ and with light neutrino mass constraints.

\begin{figure}[!hbt]
  \centering
  \includegraphics[scale=0.6]{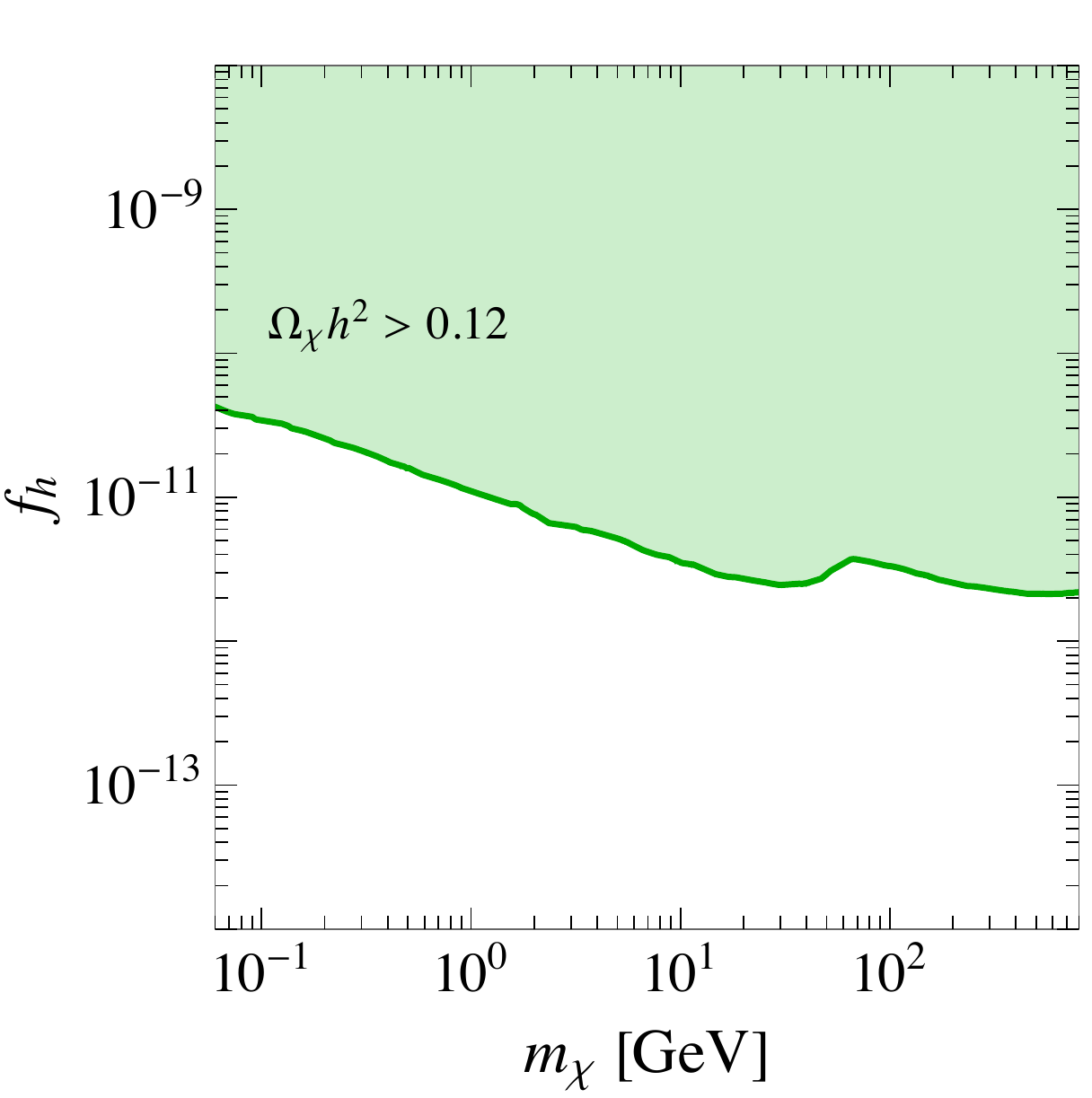}
  \caption{$f_h$ as a function of $m_\chi$. The green plain line denotes the 
contour along which $\Omega h^2 = 0.12$~\cite{Aghanim:2018eyx}. The region 
in the green shaded area is excluded as it leads to overabundant dark matter.}
  \label{fig:fh-mchi}
\end{figure}

Fig.~\ref{fig:fh-mchi} shows the values of $f_h$ required to obtain the 
observed relic density, as a function of $m_\chi$. The green plain line 
denotes the contour along which $\Omega h^2 = 0.12$~\cite{Aghanim:2018eyx}. 
Solutions in the green shaded area are excluded as they lead to overabundant 
dark matter. Values in the white area are allowed, but in this case another 
dark-matter candidate would be required.  As expected in a FIMP scenario, for 
the relic abundance of $\chi$ to meet current observations, $f_h$ must be 
small: $f_h \lesssim 3 \times 10^{-11}$ for $m_\chi \sim 0.1$ GeV.  Its 
smallness is here justified by its loop nature. Let us notice that the 
abundance of $\chi$ is insensitive to $T_R$ as long as 
$T_R \ll m_{\zeta}, m_{N_R}$. 

For the sake of example, we summarise in Table~\ref{tab:benchmarks_in} the 
values of the relevant parameters corresponding to few benchmark points 
which lead to a relic density compatible with current observations.

\begin{table}[h!]
\begin{center}
\footnotesize{
{\begin{tabular}{| c | c | c | c | c | c |}  \hline    
 $m_\chi$ [GeV] & $m_N$ [GeV] &  $m_{\zeta}$ [GeV] & $f_h$ &  $f_\chi$ & 
$\Omega h^2$  \\
 \hline   \hline 
 0.11 & $1.4 \times 10^7$ & $7.34 \times 10^4$  & $3.26 \times 10^{-11}$ & 
1 & 0.10 \\
1.13 & $1.19 \times 10^6$ & $4.47 \times 10^5$  & $1.02 \times 10^{-11}$ & 
0.24 & 0.11 \\
 2.42 &$1.6\times 10^5$ & $3.57\times 10^4$ & $7.54\times10^{-12}$ & 3.7 & 0.12\\
 21.14 & $8.3\times 10^5$ & $1.21\times 10^5$ & $2.46 \times 10^{-12}$ & 
7.19 & 0.098 \\
\hline                       
\end{tabular}}
}
\caption{Benchmark points which lead to a relic density of $\chi$ compatible 
with current observations.}
\label{tab:benchmarks_in}
\end{center}
\end{table}

\begin{figure}[!hbt]
  \centering
  \includegraphics[scale=0.5]{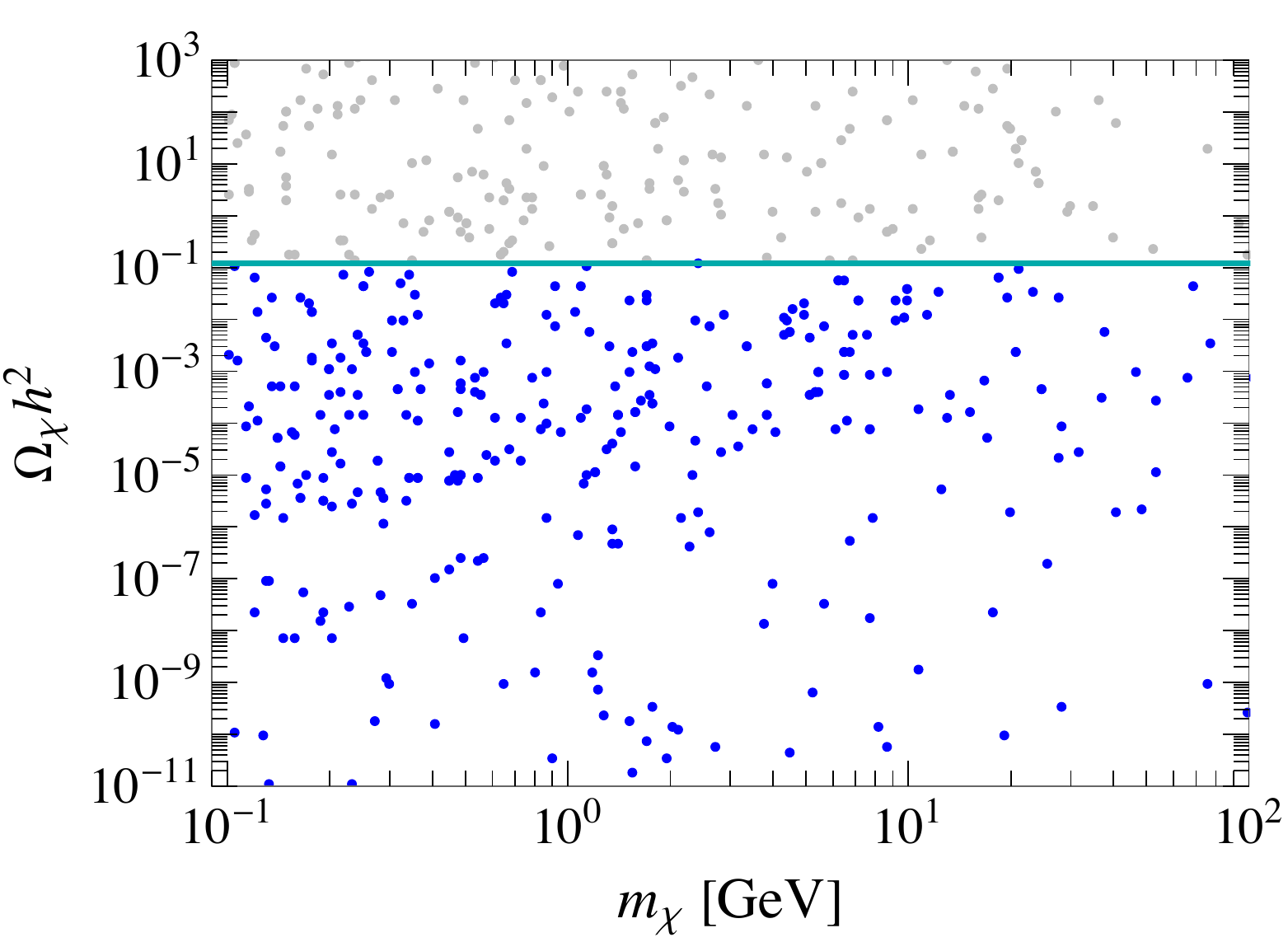}
  \caption{$\Omega h^2$ as a function of $m_\chi$. The dark cyan line denotes 
the observed value $\Omega h^2 = 0.12$ \cite{Aghanim:2018eyx}.}
  \label{fig:omega-mchi}
\end{figure}

Finally, Fig.~\ref{fig:omega-mchi} depicts the result of a numerical scan 
of the model parameter space, including the benchmark points collected in 
Table~\ref{tab:benchmarks_in}. Blue points denote solutions which are 
compatible with current cosmological observations, while grey points are 
excluded because of $\Omega h^2 > 0.12$.

\noindent \underline{\it Constraints on the Majoron}~:~
The massless singlet majoron $\theta$, which couples primarily to the light 
active neutrinos with a strength suppressed by powers of $u$, may play a role 
in cosmological and astrophysical environments. A single thermally-decoupled 
massless majoron can contribute to the effective number of neutrinos, 
$\Delta N_{\rm eff}$. This extra contribution to the radiation density depends 
on the majoron freeze-out temperature. However, because of the large values 
of the lepton number symmetry breaking scale here considered 
($u \gsim 10$ TeV) and the low reheat temperature, we find that the massless 
majoron never thermalizes with the thermal bath. Still, it could be produced 
via freeze-in through its linear coupling with the light active neutrinos. 
Also in this case, given the large values of $u$, the corresponding 
neutrino-majoron couplings lie below current cosmological constraints 
(see e.g.~\cite{Baumann:2016wac}). Finally, laboratory searches for 
non-observed lepton flavor violating rare decays 
(e.g. $\ell \rightarrow \ell^\prime \theta$) together with stellar cooling 
bounds further constrain the majoron dimensionless parameters 
$m_D^2/v u \lesssim 10^{-5}-10^{-6}$~\cite{Heeck:2019guh}. As before, 
since $u$ is much heavier than $v$ and $f_\nu$ can be quite small, the 
solutions considered in our numerical scan always fall below the current 
constraints. 

\noindent \underline{\it Concluding Remarks}~:~
It is proposed that lepton number may be a key to understanding light 
freeze-in dark matter.  In the context of spontaneously broken lepton number, 
first pointed out 40 years ago with a scalar having $L=2$, it is proposed 
that two other particles exist with nonzero lepton number, a scalar $\zeta$ 
with $L=1$ and a Majorana fermion $\chi$ with $L=2$.  It is shown that $\chi$ 
may acquire a small radiative seesaw mass and becomes freeze-in dark matter 
from the decay of the SM Higgs boson.

\noindent \underline{\it Acknowledgement}~:~
This work was supported in part by the U.~S.~Department of Energy Grant 
No. DE-SC0008541.  
VDR acknowledges financial support by the SEJI/2020/016 grant (project ``Les Fosques'') funded by
Generalitat Valenciana, by the Universitat de Val\`encia through the sub-programme “ATRACCI\'O DE TALENT 2019” and partial support by the Spanish grant FPA2017-85216-P.

\bibliographystyle{unsrt}

\end{document}